\documentclass[twocolumn,aps,prl,superscriptaddress,tightenlines]{revtex4-1}

\usepackage[dvips]{color}
\renewcommand{\section}[1]{\noindent\textbf{{#1:}}}

\newcommand{\h}{\mathcal{H}}

\newcommand{\AER}[1]{{\color{black} #1}}

	\usepackage{amsmath}
	\usepackage{amsfonts}
  	\usepackage{amssymb}
  	\usepackage{float}
	\usepackage{makeidx}
	\usepackage{amsfonts}
	\usepackage[ansinew]{}
	\usepackage[usenames,dvipsnames]{pstricks}
	\usepackage{epsfig}
    
\usepackage{xcolor}

\newcommand{\be}{\begin{equation}}
\newcommand{\ee}{\end{equation}}
\newcommand{\bea}{\begin{eqnarray}}
\newcommand{\eea}{\end{eqnarray}}

\makeindex

\begin{document}

\title{Effective speed of gravitational waves}
\author{Antonio Enea Romano}
\affiliation{Instituto de F\'isica, Universidad de Antioquia, A.A.1226, Medell\'in, Colombia}
\affiliation{ICRANet, Piazza della Repubblica 10, I--65122 Pescara, Italy}

\date{\today}

\begin{abstract}
We derive an effective equation and action for the propagation of gravitational waves (GW),  encoding the effects of  interaction and self-interaction in a time, frequency and polarization dependent effective speed. In terms of an appropriately defined effective metric, the effective action takes the form a massless Klein-Gordon action.

This effective approach predicts that for theories with matter coupled to the Einstein frame metric the ratio between gravitational and electromagnetic (EM) luminosity distance depends on the effective speed, while for Jordan frame matter coupling  it depends on the effective  Planck mass.
We discuss how the frequency and polarization dependence of the GW-EM distance ratio provides a new test of general relativity and its modifications, and more in general of the interaction of GWs with other fields.
\end{abstract}

\pacs{Valid PACS appear here}
\maketitle


\section{Introduction}
The detection of gravitational waves \cite{LIGOScientific:2016aoc} by the Laser Interferometer Gravitational Wave Observatory (LIGO) and Virgo has opened a new window on the Universe. While these observations are compatible with general relativity predictions, they also allow to put  constraints on modified gravity theories (MGT) and dark energy models. The observation of electromagnetic counterparts \cite{LIGOScientific:2017vwq} is particularly useful to test MGTs \cite{Baker:2017hug,Creminelli:2017sry,Sakstein:2017xjx,Ezquiaga:2017ekz,Wang:2017rpx}, since they allow to set stringent constraints on the difference between the speed of light $c$ and the GW speed $c_T$.

In order to establish model independent constraints on dark energy models and MGTs it can be useful to introduce a model independent equation based on using an effective stress-energy-momentum tensor approach, \cite{Romano:2018frb,Romano:2023uwf}.
This approach consists in encoding in an appropriately defined sound speed the effects due to the source terms in the perturbation equations, associated to the interaction with other fields or self-interaction. It has been successfully applied to multi-fields systems \cite{Romano:2020kmj} and MGTs \cite{Vallejo-Pena:2019hgv}, giving the correct definition of the momentum dependent sound speed and friction term able to account for the effects of the entropy.
In this paper we show that it can also be applied to GWs, allowing to find the model independent relation between the friction term and effective GW speed. 

For any theory whose field equations can be put in an "Einstein-like" form we derive a general model independent equation for GW propagation in a Friedman-Robertson-Walker (FRW) Universe, based on appropriately defining  an effective speed  encoding the effects of the source term associated to the anisotropic part of the  EST. This effective description also involves the modification of the friction term, which induces a frequency dependent difference between electromagnetic waves and GW luminosity distance $d(z)^{GW}_L$ and $d^{EM}_L(z)$.

Previous calculations of the GW propagation equation  for MGTs \cite{Kobayashi:2011nu,DeFelice:2011bh} based on the quadratic action were not including  the momentum and polarization dependency of the effective GW speed, because they are due to source terms which arise only at higher order, such as in the cubic action. 
The effective approach is in agreement and generalizes results obtained in the effective fields theory  of dark energy \cite{Gleyzes:2013ooa,Creminelli:2014wna}.
The momentum dependency of the GW effective speed is consistent with including higher order terms in the EFT of dark energy \cite{deRham:2018red,Baker:2022eiz,Creminelli:2018xsv}.

\section{Perturbations of the field equations}
Let's consider dark energy models or MGTs for which the variation of the action with respect to the metric gives  'Einstein like' field equations 

\be
G_{\mu\nu}=T^{DE}_{\mu\nu}+T^{mat}_{\mu\nu}=T_{\mu\nu}^{eff} \label{EI}\,,
\ee
where $G_{\mu\nu}$ is the Einstein tensor,  $T^{DE}_{\mu\nu}$ and $T^{mat}_{\mu\nu}$ are respectively the effective dark energy and matter ESTs, and we are using units in which $8\pi G=c=1$.
In this notation the effective Planck mass which can arise in MGTs is included in the $T_{\mu\nu}^{eff}$.
Eq.(\ref{EI}) is true for a large class of theories, for example for Horndeski's theories \cite{Horndeski:1974wa} and nonlocal theories \cite{Jaccard:2013gla}.
The calculation of $T_{\mu\nu}^{eff}$ can be cumbersome, and for this reason an action approach is often preferred, but if we treat its components as effective quantities we can  derive model independent equations.   

The advantage of perturbing the field equations is that the perturbations of $T_{\mu\nu}^{eff}$ can be treated as effective model independent quantities, \AER{under the  assumption that they are compatible with the structure of the Einstein tensor, and satisfy the energy conditions and the conservation equations implied by the Bianchi identities.}
Since the field equations have the same structure that in general relativity, the field equations of these theories can be written in a "Einstein-like" form, and this must apply also to perturbations. 

\AER{The scalar-vector-tensor (SVT) decomposition for metric and EST perturbations allows to derive model independent equations valid for an arbitrary number of fields, since the SVT decomposition is purely geometrical, and applies to any metric and EST.}

For the perturbed field equations $\delta G_{\mu\nu}=\delta T_{\mu\nu}^{eff}$ the SVT decomposition gives \cite{Kodama:1984ziu}
\be
h_A''+2 \h h_A'+ \nabla^2 h_A= a^2 \Pi^{eff}_A\label{hEI} \,,
\ee
where $A$ stands for the polarizations, the prime denotes the derivative with respect to conformal time $\eta$, $a(\eta)$ is the scale factor, $\h=a'/a$, and $\Pi_A$ is related to the traceless transverse anisotropic part of $T_{\mu\nu}^{eff}$. 

The l.h.s. of the above equation is obtained by expanding the Einstein tensor to first order in perturbations, and for this reason we get a linear differential operator, while the r.h.s. is not obtained by a perturbative expansion, but by treating the components of $\delta T_{\mu\nu}^{eff}$ as effective quantities, respecting the structure of the Einstein equations. 
This implies that the source term in eq.(\ref{hEI}) is effectively including higher order terms in perturbations, which would require to go beyond the quadratic action.
This point will be very important when comparing to previous results.




\section{Effective speed of gravitational waves}
\AER{
For comoving curvature perturbations it has been shown \cite{Romano:2018frb,Romano:2023uwf} that is possible to introduce an effective sound speed encoding the effects of entropy perturbations in multi-fields systems \cite{Romano:2020kmj}, and in MGTs \cite{Vallejo-Pena:2019hgv}.
We will show that a similar effective speed can be defined for GWs. 
Let's assume that $\hat{h}_A$ is a solution of eq.(\ref{hEI}).  We can rewrite the equation as
\be
\frac{(\hat{h}'_A a^2)'}{a^2}-\frac{(\hat{g}_A a^2)'}{a^2}- \nabla^2 \hat{h}_A=0 \label{gh} \,,
\ee
where we have defined
\be
\hat{g}_A=\frac{1}{a^2}\int a^4 \hat{\Pi}^{eff}_A\,d\eta\,, 
\ee
and the hat denotes quantities obtained by substituting the solutions of the equations of motion. Note that $\Pi_A$ is related to the coupling of GWs with other fields, so in general a system of coupled differential equations has to be solved to obtain $\hat{\Pi}_A$.

After introducing the space dependent effective GW speed $c_{T,A}$ (SEGS)  and the effective scale factor $\alpha_A$
\bea
c_{T,A}^2(\eta,x^i)=\Big(1-\frac{\hat{g}_A}{\hat{h}'_A}\Big)^{-1} &,&  \alpha_A=\frac{a}{c_{T,A}}\,, \label{ct}
\eea 
we can rewrite eq.(\ref{gh}) as
\be
\frac{1}{a^2}(\alpha_A^2 \hat{h}'_A)'-\nabla^2 \hat{h}_A=0\,. \label{zh}
\ee
Note that the effective speed $c_{T,A}$ for a given physical system is obtained by substituting the solutions of the equations of motion, and as such it is a function of space and time, not a functional of $h_A$. Taking this into account from eq.(\ref{zh}) we  obtain 
\bea
\hat{h}_A''+2 \frac{\alpha_A'}{\alpha_A} \hat{h}'_A-c_{T,A}^2 \nabla^2 \hat{h}_A&=&0 \,,\label{halpha}
\eea
showing that the SEGS is the correct definition of effective speed, since it is the coefficient of the Laplacian.

We have shown that any solution of eq.(\ref{hEI})  can be obtained by solving  eq.(\ref{halpha}) with an appropriately chosen $c_{T,A}$, but the opposite is not guaranteed, i.e. not all solutions of the effective equation (\ref{halpha}) are also solutions of eq.(\ref{hEI}).
In fact the source term $\Pi_A$ is not arbitrary, since it is related to the EST, which has to satisfy the energy and conservation conditions,  and consequently not all forms of $c_{T,A}$ are physically acceptable. These constraints on $c_{T,A}$ will be studied in a separate work.
Eq.(\ref{halpha}) can be used as a general model independent equation, which  can be used for phenomenological analysis, treating $c_{T,A}$ as a function to be determined by analyzing observational data. 


Using eq.(\ref{ct}) and dropping the hat,  we obtain 
\bea
h_A''+2  \h\Big(1-\frac{c_{T,A}'}{\h c_{T,A}}\Big) h'_A-c_{T,A}^2 \nabla^2 h_A&=&0  \,,\label{hct}
\eea
 showing that the friction term cannot be modified unless the the effective speed is time dependent.
We have dropped the hat because for phenomenological applications $h_A$ in eq.(\ref{hct}) is an unknown function which can be determined by solving the effective equation for different forms of the effective speed $c_{T,A}$.
For a given model the effective speed is given by eq.(\ref{ct}), obtained by substituting the solutions of the equations of motion, but for model independent phenomenological applications $c_{T,A}$ is a free function to be constrained by observational data. }


The advantage of the EST approach is that it allows to derive a general equation for GWs, valid for any dark energy model or MGT, including the effects of higher order terms of the anisotropic part of the EST. This is particularly useful for phenomenological studies, since it gives a general equation which can be used for  model independent analyses.

\section{Momentum effective speed}
In momentum space eq.(\ref{hEI}) takes the form
\be
\tilde{h}_{A}''+2 \h \tilde{h}_{A}'+ k^2 \tilde{h}_{A}= a^2 \tilde{\Pi}^{eff}_{A}\,,
\ee
where we are denoting with $\tilde{h}_A(\eta,k)$ and $\tilde{\Pi}_A(\eta,k)$  the Fourier transform of $h_A$ and $\Pi_A$. 
Following a similar procedure to the one used to derive eq.(\ref{hct}) we obtain 
\bea
\tilde{h}_{A}''+2 \frac{\tilde{\alpha_A}'}{\tilde{\alpha_A}} \tilde{h}_{A}'+\tilde{c}^2_{T,A} k^2 \tilde{h}_{A}&=&0 \,,\\ \label{halphak}
\tilde{h}''_{A}+2 \h\Big(1-\frac{\tilde{c}'_{T,A}}{\h \tilde{c}_{T,A}}\Big) \tilde{h}'_{A}+\tilde{c}^2_{T,A} k^2 \tilde{h}_{A}&=&0  \,,\label{hkct}
\eea
where we have defined the momentum effective GW speed (MEGS) according to

\bea
\tilde{c}_{T,A}^2(\eta,k)&=&\Big(1-\frac{\tilde{g}_{A}}{\tilde{h}'_{A}}\Big)^{-1} \,,\\ \tilde{g}_{A}&=&\frac{1}{a^2}\int a^4 \tilde{\Pi}_{A}^{eff}\,d\eta\,, \label{mess}
\eea 
and  $\tilde{\alpha}_A^2=a^2/\tilde{c}_{T,A}^2$.
Note that the MEGS $\tilde{c}_T(\eta,k)$ defined in eq.(\ref{mess})  is not the Fourier transform of the SEGS $c_T(\eta,x^i)$ defined in eq.(\ref{ct}), since the Fourier transform of the terms $c_T\nabla^2 h_A$ and $h'_A c_T'/c_T$ in eq.(\ref{hct}) are a convolution, not a product, of their transform.
The definition of MEGS is mathematically convenient since it involves a minimal modification of  eq.(\ref{hEI}). 

\section{Frequency dependency of the GW luminosity distance}
In general relativity the amplitude of GWs is inversely proportional to the electromagnetic  luminosity distance $d^{EM}_L(z)$, but a modification of the friction term of the GW propagation equation induces a difference   \cite{Belgacem:2017ihm} with respect to the GW luminosity distance $d_L^{GW}$. In order to find the effects on the luminosity distance it is convenient to rewrite eq.(\ref{hkct}) as
\be
\tilde{\chi}_A''+\Big(\tilde{c}_{T,A} k^2 -\frac{\tilde{\alpha}_A''}{\tilde{\alpha}_A}\Big)\tilde{\chi}_A=0 \,,
\ee
where we have defined $\tilde{h}_A=\tilde{\chi}_A/\tilde{\alpha}_A$. \AER{On sub-horizon scale $\alpha''/\alpha$ can be neglected, and since the amplitude $\tilde{h}_A$ is proportional to $1/\tilde{\alpha}$ instead of $1/a$, we get that 
\be
\frac{d_{L}^{GW}}{d_{L}^{EM}}(z)=\frac{a(z)}{\tilde{\alpha}_A(z)}\frac{\tilde{\alpha}_A(0)}{a(0)}=\frac{\tilde{c}_{T,A}(z,k)}{\tilde{c}_{T,A}(0,k)}\,,\label{dgw}
\ee
where we have defined $\tilde{\alpha}_A(z)=a(z)/\tilde{c}_{T,A}(\eta(z),k)=a(z)/\tilde{c}_{T,A}(z)$, and used $d_L^{GW}=r\,\tilde{\alpha}(0)/\tilde{\alpha}(z)$, $d_L^{EM}=r\, a(0)/a (z)$, assuming $(1+z)=a(0)/a(z)$, i.e. that  matter is minimally coupled to the Einstein frame metric. Eq.(\ref{dgw}) generalizes the result obtained in \cite{Romano:2023ozy} using the EFT.}
In general relativity $\tilde{c}_{T,A}=1$, $\tilde{\alpha}(z)=a(z)$, and  we recover the well known result that $d_L^{GW}(z)=d_L^{EM}(z)$. Neglecting higher order interaction effects the effective speed  $\tilde{c}_{T,A}$ reduces to $c_T$, and we recover the results obtained in \cite{Romano:2023ozy} using the EFT of dark energy.

\section{Effective Lagrangian and metric}
The Lagrangian density for which the Lagrange's equation gives eq.(\ref{zh}) is
\be
\mathcal{L}^{eff}_h=\frac{a^2}{c_{T,A}^2}\Big[ h'^2_A-c_{T,A}^2 (\nabla h_A)^2\Big],\label{L}
\ee
in agreement, and generalizing, the EFT calculations in the Einstein frame \cite{Creminelli:2014wna}.

The effective Lagrangian  can be obtained from the general relativity Lagrangian density

\be
\mathcal{L}_h^{GR}=a^2\Big[h'^2_A-c^2 (\nabla h_A)^2\Big]=\sqrt{-g}(\partial_{\mu}h_A \partial^{\mu}h_A)\,,
\ee
via the transformation
\be
a\rightarrow \alpha_A=\frac{a}{c_{T,A}} \quad,\quad c\rightarrow c_{T,A} \,,\label{trans}
\ee
where we have denoted with $c$ the speed of light, to avoid ambiguity.
This in agreement with eq.(\ref{halpha}), which shows that $\alpha_A$ can be regarded as an effective scale factor.
The effective action  can be written as
\be
\mathcal{L}^{eff}_h=\sqrt{-g_A}(\partial_{\mu}h_A \partial^{\mu}h_A) \,,
\ee
in terms of the effective metric
\be
ds^2_A=a^2\Big[c_{T,A}d\eta^2-\frac{\delta_{ij}}{{c_{T,A}}}dx^idx^j\Big] \label{geff} \,,
\ee
for which the wave eq.(\ref{hct}) can be written in terms of the covariant d'Alembert operator as
\be
\square h_A=\frac{1}{\sqrt{-g_A}}\partial_{\mu}(\sqrt{-g_A}\partial^{\mu}h_A)=0 \,.
\ee
This equation shows that the effects of the interaction of the graviton with other fields can be effectively described as the propagation in vacuum through a space with the effective metric given in eq.(\ref{geff}). Note that the above effective  metric is only valid to describe GWs propagation, it is not a modification of the background metric describing the full cosmological model. For scalar perturbations for example a different effective metric and action can be defined in terms of the momentum effective sound speed (MESS) \cite{Romano:2018frb,Romano:2023uwf}.

In the action approach the EST on the r.h.s. of the Einstein equations originates from the interaction of tensor perturbations with themselves or other fields. Based on this we can obtain the effective action in eq.(\ref{L}) by introducing  higher order   interaction terms in the Einstein frame Lagrangian as $\mathcal{L}(h_A,\phi^i)$

\be
\mathcal{L}_h=a^2\Big[h'^2_A-(\nabla h_A)^2+\mathcal{L}(h_A,\phi^i)\Big] \,\label{LEint}\\
\ee
which can be rewritten in the form of eq.(\ref{L})
by introducing the effective GW speed as
\be
c^2_{T,A}=\Big(1+\frac{\mathcal{L}}{h'^2_A}\Big)^{-1} \label{ctS}\,,
\ee
and $\phi^i$ denotes abstractly all the other fields the graviton is coupled to, including itself, or another polarization. Note that effective speed is obtained by substituting in eq.(\ref{ctS}) the solutions of the equations of motion, when possible \cite{Pons:2009ch}.
In the effective field theory of inflation \cite{Noumi:2014zqa,Creminelli:2014wna}, denoting with $\delta K_{\mu\nu}$ the perturbation of
the extrinsic curvature of the spatial slices, the leading order term for $\mathcal{L}$  is 
\be 
\mathcal{L}^{(2)}\propto  \bar{M}_2(\eta) \delta K_{\mu\nu}\delta K^{\mu\nu}\propto \bar{M}_2(\eta) h'^2_A,
\ee
which induces a time dependent $c_T(\eta)$, and an action in agreement with eq.(\ref{L}). Higher order terms $\mathcal{L}^{(i>2)}$ can induce a space and polarization dependent $c_{T,A}(\eta,x)$, which encodes the effects of the higher order interaction terms \cite{Note3}.
By comparing eq.(\ref{ct}) and  eq.(\ref{ctS}) we obtain the relation between the interaction  Lagrangian $\mathcal{L}$ and the anisotropic part of the EST  $\mathcal{L}=g^2_A-2h'_A g_A$, which allows to relate the effective action approach with the EST.

\section{Jordan frame formulation}
Eq.(\ref{L}) should be valid for any system satisfying eq.(\ref{EI}), and in particular \AER{it can be shown that it is consistent with the Horndeski's theory Lagrangian density in Jordan frame \cite{Kobayashi:2011nu}

\be
\mathcal{L}^{Hor}_{h,J}=\tilde{a}^2 M_*^2 \Big[ h_A'^2-c_T^2(\nabla h_A)^2\Big]\,, \label{Lh}
\ee
where we are denoting with $\tilde{a}$ the Jordan frame scale factor.
The Lagrangian density in eq.(\ref{L}) is in the Einstein frame, since it gives the equations which were derived from the "Einstein like" eq.(\ref{EI}), while eq.(\ref{Lh}) is in the Jordan frame.
In order to find the transformation between the two frames we can note that the non perturbed field equations for the Horndeski's theory can be written in the Jordan frame as \cite{Kobayashi:2011nu}  
\bea
\Omega^2 G_{\mu\nu}^J=T^{eff,J}_{\mu\nu} \label{EJ}&,&
\Omega^2=G_4+\frac{1}{2}\nabla_{\lambda}G_5\nabla^{\lambda}\phi\,,
\eea 
from which we can find the conformal transformation from the Jordan to the Einstein frame, $g_{\mu\nu}=\Omega^2 \tilde{g}_{\mu\nu}$, with $\Omega=M_* \,c_T$, in agreement with applications of the EFT of dark energy   \cite{Kennedy:2017sof}.  
The Einstein frame effective Lagrangian can be converted to the Jordan frame by the conformal transformation $a=\Omega \, \tilde{a}$ 
\bea
\mathcal{L}^{eff}_{h,J}&=&\frac{\tilde{a}^2\Omega^2}{c^2_{T,A}}\Big[h'^2_A-c_{T,A}^2(\nabla h_A)^2\Big]= \nonumber \\
&=&\tilde{a}^2 M_A^2 \Big[ h_A'^2-c^2_{T,A}(\nabla h_A)^2\Big]\label{LJ2} \,,
\eea
where $M_A(\eta,x_i)$ is the space and polarization dependent effective Planck mass in the Jordan frame, defined as
\bea
M_A=\frac{\Omega}{c_{T,A}}=\frac{M_*}{1+\delta_A} &,& c_{T,A}=c_T(1+\delta_A) \,.
\eea
The above action is in agreement with, and generalizes, the results obtained in the EFT of dark energy in the Jordan frame \cite{Gleyzes:2013ooa}.
The corresponding GW propagation equation is
\be
h_A''+2  \tilde{\h}\Big(1+\frac{M_A'}{\tilde{\h} M_A}\Big) h'_A-c_{T,A}^2 \nabla^2 h_A=0  \,.\label{hJ}
\ee
The above equation is consistent  with eq.(\ref{halpha}) with $\alpha_A=\tilde{a} M_A$. 
At leading order, i.e. ignoring higher order interaction terms,  we recover the commonly used \cite{Lagos:2019kds} Jordan frame action and equation. 

Analogously to the Einstein frame eq.(\ref{LEint}), eq.(\ref{LJ2}) could also be obtained from the Lagrangian 
\be
\mathcal{L}_J=\tilde{a}^2 M^2_* \Big[ h_A'^2-c^2_T(\nabla h_A)^2+\mathcal{L}_J^{i>2}\Big] \label{LJ0} \,,
\ee
by defining 
\bea 
1+\delta_A&=&\Big(1+\frac{\mathcal{L}_J^{i>2}}{h_A'^2}\Big)^{-1/2}, 
\eea
where $\mathcal{L}_J^{i>2}$
denotes the Jordan frame higher order interaction terms. The effective metric in the Jordan frame is the conformal transformation of that in the Einstein frame given in eq.(\ref{geff})
\be
ds_{J,A}^2=\Omega^2 ds_{E,A}^2\,.
\ee
}

\section{Momentum space effective Lagrangian}
In momentum space the effective Lagrangian density is
\be
\mathcal{L}^{eff}_{h,k}=\tilde{\alpha}_A^2\Big[\tilde{h}'^2_A+  \tilde{c}^2_{T,A}k^2 \tilde{h}_{A}\Big]\,,
\ee
which can be obtained from the general relativity action
\be
\mathcal{L}_{h,k}^{GR}=a^2\Big[\tilde{h}'^2_A+ c^2 k^2 \tilde{h}_{A}\Big]\,,
\ee
\AER{
by the transformation 
\bea
a\rightarrow \tilde{\alpha}_A=\frac{a}{\tilde{c}_{T,A}}=\tilde{a}\tilde{M}_A &,&c\rightarrow \tilde{c}_{T,A} \,,\label{transk} \\
\tilde{c}_{T,A}=c_T(1+\tilde{\delta}_A) &,& \tilde{M}_A=\frac{M_*}{1+\tilde{\delta}_A} \,,
\eea
where $\tilde{M}_A(\eta,k)=\Omega/\tilde{c}_{T,A}$ is the momentum and polarization dependent effective Planck mass and
\bea
(1+\tilde{\delta}_A)&=&\Big(1+\frac{\mathcal{\tilde{L}}_J^{i>2}}{\tilde{h}'^2_A}\Big)^{-1}\,.
\eea
}
This is in agreement with eq.(\ref{halphak}), which shows that $\tilde{\alpha}_A$ can be regarded as an effective scale factor, but in this case it is momentum dependent.
As previously mentioned, the quantities $\tilde{\alpha}_A$ and $\tilde{\delta}_A$ are not the Fourier transform of $\alpha_A$ and $\delta_A$. More details about the momentum space effective action derivation can be found in \cite{Romano:2023uwf}.

\section{Frame dependence and matter coupling}
\AER{Physical observables of a given theory should can be computed in any frame, but changing the metric to which matter fields are minimally coupled to in the Lagrangian corresponds to introducing different theories, which have different observable predictions, since it changes the scale factor redshift relation \cite{Romano:2023ozy}.}

Under a general conformal transformation, defined by $a=\Omega\, \tilde{a}$,  the Einstein frame effective Lagrangian in eq.(\ref{L}) can be written as 
\be
\frac{\Omega^2\tilde{a}^2}{c_{T,A}^2}\Big[ h'^2_A-c_{T,A}^2 (\nabla h_A)^2\Big]
\ee
giving the propagation equation 
\be
h_A''+2  \tilde{\h}\Big(1-\frac{c_{T,A}'}{\tilde{\h} c_{T,A}}+\frac{\Omega'}{\tilde{\h}\Omega}\Big) h'_A-c_{T,A}^2 \nabla^2 h_A=0  \,,\label{hctO}
\ee
where $\tilde{\h}=\tilde{a}'/\tilde{a}$,
from which we can see that while the GW speed does not depend on the conformal transformation,  the friction terms does. 

As expected, eq.(\ref{hctO}) gives eq.(\ref{hJ}) for $\Omega=c_T M_*$.
Note that, due to the conformal invariance of GWs, eq.(\ref{halpha}) is valid also in the new frame, since $\alpha_A=a/c_{T,A}=\Omega\tilde{a}/c_{T,A}=\tilde{a} M_A$, i.e. the apparent modification of the friction term has no real physical effect \cite{Romano:2023ozy} on GWs propagation. 
\AER{

An observable effect arises if matter is minimally coupled to the metric $\tilde{g}_{\mu\nu}$ instead of $g_{\mu\nu}$, in which case the redshift is given by $1+z=\tilde{a}(0)/\tilde{a}(z)$, and we get a different GW-EM distance ratio
\be
\frac{d_{L,A}^{GW}}{d_{L,A}^{EM}}(z)=\frac{\tilde{a}(z)}{\tilde{a}(0)}\frac{\tilde{\alpha}_A(0)}{\tilde{\alpha}_A(z)}=\frac{\tilde{c}_{T,A}(z)}{{\tilde{c}_{T,A}(0)}}\frac{\Omega(0)}{\Omega(z)}\,,\label{dgwO}
\ee
which in the case of the Jordan frame coupling, corresponding to $\Omega= M_* \, c_T$, gives
\be
d_{L,A}^{GW}(z)=\frac{\tilde{M}_A(0,k)}{\tilde{M}_A(z,k)}d_L^{EM}(z)\,,\label{dgwO}
\ee
generalizing the results obtained using the EFT of dark energy \cite{Romano:2023ozy}, or those obtained in some specific modified gravity theories \cite{Belgacem:2017ihm,Dalang:2019rke}. The momentum and polarization dependence is due to including the effects of higher order interaction terms neglected in leading order calculations.}

In a frame different from the Einstein frame the friction term can be modified even if $c_{T,A}$ is constant, but this is just due to the change in the Hubble parameter due to the scale factor redefinition, not to a real effect on the GW propagation, since tensor perturbations are invariant under conformal transformations \cite{Tsujikawa:2014uza}. In fact eq.(\ref{hct}) is valid in any frame, since the friction term in different frames is the same function of space and time $\alpha'_A/\alpha_A$, written in different forms according to $\alpha_A=a^2/c_{T,A}^2=\Omega^2\tilde{a}^2/c_{T,A}^2$.

\section{Observational constraints}
Eq.(\ref{dgw}) and eq.(\ref{dgwO}) allow to constrain  the effective GW speed and the effective Planck mass for different types of theories, using the luminosity distance ratio between $d_L^{GW}(z)$ and $d_L^{EM}(z)$. 
For a GW of given frequency the observation of a single electromagnetic counterpart at redshift $z_{ct}$ allows to set a single constraint on $\tilde{c}_{T,A}(z_{ct},k_{ct})$, but only observations at different frequencies and redshifts allow to fully constrain the MEGS.  Combining data from different GW  detectors, sensitive to different frequency bands, such as  LISA, Einstein Telescope and Cosmic Explorer, would be particularly useful.

For dark sirens a statistical method similar to the one employed for the estimation of the Hubble constant \cite{Gray:2021sew} can be adopted to estimate the redshift, allowing to constrain $\tilde{c}_{T,A}$ with much more data than for bright sirens, which are in fact expected to be a limited portion of the GWs detections.
The electromagnetic counterpart observation of  GW170817 \cite{LIGOScientific:2017vwq} is imposing a tight constraint on  the low redshift value of $\tilde{c}_{T,A}(z)$ at the frequencies probed by LIGO-Virgo, so a possible phenomenological ansatz for theories with Einstein frame matter coupling could be
\AER{
\be
\frac{d^{GW}_{L,A}(z)}{d^{EM}_L(z)}=1+\frac{\gamma_A(k)}{(1+z)^{n_A(k)}} \tanh\Big[\frac{k-k^*_A}{\sigma_A(k)}\Big] \,,\label{ansatz}
\ee
}
where $k^*_A$ is the scale at which a transition takes place, $\gamma_A$ controls the size of the change, and $\sigma_A$ the sharpness of the transition.
Other constraints on the $\tilde{c}_{T,A}$  can be obtained from the waveform \cite{Harry:2022zey}, since stretching or squeezing is expected, and they could be combined with those on the luminosity distance given in eq.(\ref{ansatz}).

For theories with Jordan frame coupling the distance ratio can be used to constrain the effective Planck mass, testing its possible polarization and frequency dependency. 
A possible model independent ansatz could be a a generalization of the parametrizations studied in \cite{LISACosmologyWorkingGroup:2019mwx}, for example
\AER{
\bea
\frac{d^{GW}_{L,A}(z)}{d^{EM}_L(z)}&=&\Xi_A(k)+\frac{1-\Xi_A(k)}{(1+z)^{n_A(k)}}\,.
\eea
}


\section{Conclusions}
\AER{
We have derived  effective equations and Lagrangians for the GWs propagation in the Einstein and Jordan frame, which can be used for model independent analysis of GWs observational data.
We have provided two derivations, one based on the EST, and the other on the effective action, generalizing the results obtained in the EFT of dark energy \cite{Gleyzes:2013ooa}.
The effective description is based on two  quantities: the effective GW speed and the effective Planck mass, which can both be polarization and frequency dependent, due to higher order interaction terms. In the Einstein frame the effective Lagrangian depends only on the effective GW speed.
This effective approach allows to derive model independent relations between the GW and electromagnetic luminosity distance, which can be tested observationally with dark or bright sirens.}

While this model independent approach is useful for phenomenological analyses of GWs observations, it should be noted that the  EST includes all the effects of matter and dark energy in a single object, and without other observations it would not be possible to disentangle them. For example, the observation of the difference between the luminosity distance of the two polarizations of a GW event could be caused either by the propagation through an anisotropic matter medium \cite{Romano:2023lxf}, or by a MGT. 
In order to test specific dark energy models it will be important to perform higher order perturbations calculations, in order to compute the anisotropy effects which are not included in the quadratic action, and in the EST approach are treated as effective model independent quantities.

\section{Acknowledgments}
I thank Misao Sasaki, Tessa Baker, Sergio Vallejo, Riccardo Sturani, Rogerio Rosenfeld, Claudia de Rham, Nicola Tamanini, Suvodip Mukherjee, Mairi Sakellariadou, Shinji Mukohyama and Juan Maldacena for interesting discussions. I thank the ICTP-SAIFR for the kind hospitality. AER was supported by the UDEA projects  2021-44670,  2019-28270, 2023-63330.


\bibliography{Bibliography}
\bibliographystyle{h-physrev4}

\end{document}